\documentclass[aps,twocolumn]{revtex4-1}

\usepackage[utf8]{inputenc}
\usepackage{amsmath}
\usepackage{amssymb}
\usepackage{hyperref}
\usepackage{braket}
\usepackage{graphicx}
\usepackage{xcolor}

\newcommand{\Tr}{\text{Tr}}

\begin{document}

\title{Nonlinear Phase Gates as Airy Transforms of the Wigner Function}
\author{Darren W. Moore}
\email{darren.moore@upol.cz}
\author{Radim Filip}
\email{filip@optics.upol.cz}
\affiliation{Department of Optics, Palack\'{y} University, 17. listopadu 1192/12, 771 46 Olomouc, Czech Republic}

\begin{abstract}
Low-order nonlinear phase gates allow the construction of versatile higher-order nonlinearities for bosonic systems and grant access to continuous variable quantum simulations of many unexplored aspects of nonlinear quantum dynamics. The resulting nonlinear transformations produce, even with small strength, multiple regions of negativity in the Wigner function and thus show an immediate departure from classical phase space. Towards the development of realistic, bounded versions of these gates we show that the action of a quartic-bounded cubic gate on an arbitrary multimode quantum state in phase space can be understood as an Airy transform of the Wigner function. This toolbox generalises the symplectic transformations associated with Gaussian operations and allows for the practical calculation, analysis and interpretation of explicit Wigner functions and the quantum non-Gaussian phenomena resulting from bounded nonlinear potentials.
\end{abstract}

\maketitle

\section{Introduction}

Versatile quantum simulation with bosonic systems requires a universal set of gates incorporating at least one, often experimentally demanding, nonlinear operation. One such gate set is composed of linear phase gates (alongside the Fourier transform) and at least one nonlinear phase gate, which together are central components of universal quantum information processing with continuous variables~\cite{ghose_non-gaussian_2007,miwa_demonstration_2009}. The lowest order nonlinear phase gate's effect on the Wigner function is highly nontrivial, contrasting with the simple and analytically expressible transformation of the phase space operators associated with linear phase gates. Whereas linear phase gates merely displace and shear the Wigner function, even preserving Gaussianity, a nonlinear phase gate introduces complex oscillations, sub-Planck structures and negativity~\cite{dragt_how_1998}. Such features appear as the outcome of quantum interference between the larger positive phase space structures associated with the classical approximation to the dynamics. This quantum interference, also associated with the nonlinear dynamics of fully continuous variable systems, creates Wigner functions fundamentally different from those generated by finite polynomials modulated by a Gaussian envelope such as Fock states or finite superpositions of them~\cite{groenewold_principles_1946}, or the discrete and localised quantum interference of Gaussian states represented by the cat states~\cite{szabo_construction_1996}, compass states~\cite{zurek_sub-planck_2001} or even GKP states~\cite{gottesman_encoding_2001}. These states are typically inaccessible via unitary processes acting on localised (e.g. Gaussian) states. The quantum non-Gaussian states resulting from continuously nonlinear dynamics qualitatively show shallower and interference fringes that are more broadly spread throughout phase space.

The classification of such states faces a further complication once non-pure states are involved. While pure states possess positive Gaussian Wigner functions, and are otherwise negative~\cite{hudson_when_1974,littlejohn_semiclassical_1986}, the classification breaks down for mixed states~\cite{mandilara_extending_2009,van_herstraeten_quantum_2021}. Instead, the set of Wigner positive states is found to be a proper superset of the convex hull of the Gaussian states, so that there are mixed non-Gaussian states with positive Wigner functions~\cite{filip_detecting_2011}. Therefore, understanding nonlinear gate operations on mixed states in terms of the Wigner function is essential. The method we describe below is independent of the purity of the initial state and thus facilitates investigations into such operations.

The prototypical nonlinear phase gate is the cubic phase gate, which comes with an associated set of cubic phase states resulting from its application to Gaussian states. It was recognised fairly early that the Wigner function of the cubic phase gate acting on the unphysical momentum eigenstate is in fact an Airy function of the canonical variables~\cite{ghose_non-gaussian_2007}. This continues to be the case when the unphysical momentum state is replaced by the physical harmonic oscillator ground state~\cite{brunelli_linear_2019} or even by an arbitrary Gaussian state~\cite{riera-campeny_wigner_2024}. However the cubic nonlinearity is unbounded from below, which provides another possible source of unphysicality~\cite{felicetti_universal_2020,minganti_non-gaussian_2023} or may provide resources inaccessible to lower-bounded Hamiltonians. In order to compensate for this the Hamiltonian must be bounded by a higher-order nonlinearity~\cite{felicetti_universal_2020,minganti_non-gaussian_2023}.
In this article we provide an analytic description of the effect of cubic or quartic phase gates acting on an arbitrary density operator in phase space in terms of Airy transforms of the Wigner function. This more general result allows universal gate sets for quantum computation to be in principle implemented fully analytically. Additionally, towards a deeper understanding of continuous nonlinear dynamics in the large mass regime (or, equivalently, in the short time regime), the combined effect of these gates allows examination of the quartic-bounded cubic phase gates and their critical comparison to unbounded cubic gates and tilted double-well gates. This opens a new road to investigate and simulate highly nonclassical phenomena through the Wigner function.

\section{Results}
\subsection{Airy Transforms of the Wigner Function}

The Wigner function in phase space is the best candidate to represent the nonclassical and quantum non-Gaussian aspects of states resulting from nonlinear dynamics in systems of bosonic continuous variables. Significantly, individual points of the Wigner function $W(q,p)=\frac{1}{\pi\hbar}\Tr\left(D(q,p)\Pi D^\dagger(q,p)\rho\right)$, of the quantum state $\rho$, are directly measurable via interferometry via the parity operator $\Pi=\int dx\ket{-x}\bra{x}=(-1)^{a^\dagger a}$, where $a$ is the bosonic annihilation operator~\cite{laiho_probing_2010,eriksson_universal_2023}. The displacement $D(q,p)=e^{\frac{1}{\sqrt{2}\hbar}\left((q+ip)a^\dagger-(q-ip)a\right)}$ scans over the phase space, similar to other interference experiments, to extract information on phase space superpositions. The Wigner function can be reformulated~\cite{royer_wigner_1977} as the Fourier transform of the anti-diagonal of the density operator
\begin{equation}\label{Wigner}
    W(q,p)=\frac{1}{\pi\hbar}\int~e^{\frac{2ipt}{\hbar}}\braket{q-t|\rho|q+t}~dt\,,
\end{equation}
expressed in the coordinate basis corresponding to $\hat{q}$~\cite{weinbub_recent_2018}. Note that this defines a convention for the Fourier transform, which will be adopted later in the discussion of Airy transforms. It follows that the inverse transform produces the anti-diagonal as a function of $t$ from the Wigner function, that is
\begin{equation}
    \braket{q-t|\rho|q+t}=\int e^{-\frac{2ipt}{\hbar}}W(q,p)~dp
\end{equation}
As is well known, the Wigner function forms a quasi-probability distribution for the phase space variables $q$ and $p$, corresponding to the canonical operators satisfying the commutation relation $[\hat{q},\hat{p}]=i\hbar$.

Unitary transformations of $\rho$, whether representing dynamics or quantum gates, can then be interpreted directly as transformations of the Wigner function. Unitary transformations that are bilinear in the operators $q$ and $p$, called Gaussian unitaries, correspond to linear symplectic maps $\mathcal{S}$ of the phase space variables for the Wigner function~\cite{weedbrook_gaussian_2012}. The Wigner function is transformed by the applying the corresponding symplectic matrix $S$ to these variables. Collecting the phase space variables into a vector $\mathbf{x}=\begin{pmatrix}q \\p\end{pmatrix}$, we write
\begin{equation}
    W(\mathbf{x})\rightarrow W(S\mathbf{x})\,.
\end{equation}
In particular, the phase gates take the operator form $U_n=e^{-i\frac{\gamma}{n\hbar}\hat{q}^n}$ and implement the following unitary transformations on the quadrature operators
\begin{align}
    U_n\hat qU_n^\dagger&=\hat q\\
    U_n\hat pU_n^\dagger&=\hat p+\gamma \hat q^{n-1}\equiv\mathcal{S}_n(\hat p)\,.
\end{align}
Note that for later notational simplicity the ordering of the unitary operators is reversed compared to normal time evolution. By applying these phase gates directly to $\rho$ in Eq.~(\ref{Wigner}) the Wigner function is transformed into 
\begin{align}\label{WigPhaseGates}
    W(q,p)=&\frac{1}{\pi\hbar}\int~e^{\frac{2ipt}{\hbar}}\braket{q-t|U_n\rho U_n^\dagger|q+t}~dt\\
    =&\frac{1}{\pi\hbar}\int~e^{\frac{2ipt}{\hbar}}e^{-i\frac{\gamma}{n\hbar}\left((q-t)^n-(q+t)^n\right)}\braket{q-t|\rho|q+t}~dt\,.
\end{align}
This amounts to the addition of an extra phase term in the Wigner function integral. For $n=1,2$ the new exponential term is linear in $t$, and this amounts to a relabelling $p\rightarrow p+\gamma q^{n-1}$, corresponding to the symplectic map $\mathcal{S}_n$. Then the integral is still interpreted as a Wigner function, with a linear transformation $\mathcal{S}_n(p)$ of the momentum variable with respect to the original. That is, $W(q,p)\xrightarrow{U_n}W(q,\mathcal{S}_n(p))$ for $n=1,2$. However, for nonlinear phase gates ($n>2$) the transformation of the Wigner function does not correspond to the nonlinear transformation $\mathcal{S}_n(p)$ for the phase space variables of the Wigner function. Furthermore attempting to apply the transformation in this way does not recover the Liouvillian density in the classical limit~\cite{dragt_how_1998}.

To illustrate this difference concretely, consider the case $n=3$ where an extra exponential term appears which is not linear in the integration variable $t$. That is,
\begin{align}\label{cubphas}
    W(q,p)&\xrightarrow{U_3}\frac{1}{\pi\hbar}\int~e^{i\frac{2\gamma}{3\hbar}t^3}e^{\frac{2i(p+\gamma q^2)t}{\hbar}}\braket{q-t|\rho|q+t}~dt\\
    &\ne W(q,\mathcal{S}_3(p))\,.\nonumber
\end{align}
The extra phase $e^{i\frac{2\gamma}{3\hbar}t^3}$ prevents the interpretation of the integral as a Wigner function with a simple transformation of the phase space variables, unlike the case of Gaussian unitaries. Remarkably even in the case $n=4$ the extra exponential term involves only the cube of the integration variable.

Indeed, for both $n=3,4$ it is possible to interpret this integral as the Airy transform of the original Wigner function, along with the nonlinear transformation of the momentum variable $\mathcal{S}_n(p)$. Let us first introduce the Airy transform.

The Airy transform~\cite{widder_airy_1979,vallee_airy_2010} is defined as the convolution product of a function $f(x)$ with the family of Airy functions $\text{Ai}(x;\alpha)=\frac{1}{2\pi|\alpha|}\int e^{i\left(\frac{z^3}{3}+\frac{xz}{\alpha}\right)}dz$. For our purpose it is useful to let $\alpha\rightarrow\frac{\hbar}{2}\alpha$ so that $\text{Ai}(x;\alpha)=\frac{1}{\pi\hbar|\alpha|}\int e^{i\left(\frac{z^3}{3}+\frac{2xz}{\hbar\alpha}\right)}dz$ and we may write the convolution product explicitly as
\begin{align}
    \mathcal{A}_\alpha[f](x)&=f*\text{Ai}(x;\alpha)\\
    &=\frac{1}{\pi\hbar|\alpha|}\int\int e^{i\left(\frac{z^3}{3}+\frac{2(x-\tau)z}{\hbar\alpha}\right)}f(\tau)dzd\tau\\
    &=\frac{1}{\pi\hbar|\alpha|}\int e^{i\left(\frac{z^3}{3}+\frac{2xz}{\hbar\alpha}\right)}\hat{f}\left(\frac{z}{\alpha}\right)dz\,,
\end{align}
where $\hat f$ is the inverse Fourier transform using the aforementioned Wigner function convention.

We now return to the effect of the nonlinear phase gates $U_3$ and $U_4$ in phase space [see Eq.~(\ref{WigPhaseGates})]. It will be useful for what follows to make the substitution $t=\frac{z}{\alpha}$ with $\alpha\ne0\in\mathbb{R}$, where we now write explicitly
\begin{align}
    W(q,p)&\xrightarrow{U_3}\frac{1}{\pi\hbar|\alpha|}\int~e^{\frac{2i\mathcal{S}_3(p)z}{\alpha\hbar}}e^{i\frac{2\gamma}{3\alpha^3\hbar}z^3}\braket{q-\frac{z}{\alpha}|\rho|q+\frac{z}{\alpha}}~dz\nonumber\\
    W(q,p)&\xrightarrow{U_4}\frac{1}{\pi\hbar|\alpha|}\int~e^{\frac{2i\mathcal{S}_4(p)z}{\alpha\hbar}}e^{i\frac{2\gamma q}{\alpha^3\hbar}z^3}\braket{q-\frac{z}{\alpha}|\rho|q+\frac{z}{\alpha}}~dz\,,\label{WignerAiry}
\end{align}
and we have used the symplectic map $\mathcal{S}_n$ notation introduced earlier. Note that if $\alpha<0$ then the integration limits swap in the sense $\int_{-\infty}^{\infty} \rightarrow\int_\infty^{-\infty}$. To return to the standard order a minus sign is factored out, resulting in the absolute value $|\alpha|$.

To interpret these transformations as Airy transforms, consider the Airy transform of the Wigner function $W(q,p)$ with respect to $\mathcal{S}_n(p)$. We may write this as
\begin{align}
    \mathcal{A}_\alpha[W](\mathcal{S}_n(p))&=\int e^{i\left(\frac{z^3}{3}+\frac{2\mathcal{S}_n(p)z}{\hbar\alpha}\right)}\int e^{-i\frac{2\tau z}{\hbar\alpha}}W(q,\tau)\frac{dzd\tau}{\pi\hbar|\alpha|}\\
    =\frac{1}{\pi\hbar|\alpha|}&\int e^{i\left(\frac{z^3}{3}+\frac{2\mathcal{S}_n(p)z}{\hbar\alpha}\right)}\braket{q+\frac{z}{\alpha}|\rho|q-\frac{z}{\alpha}}dz\,,
\end{align}
where we have used the inverse Fourier transform to retrieve the anti-diagonal of the density operator from the Wigner function. It is immediate that an appropriate choice of $\alpha$ in the Wigner functions of Eqs.~\ref{WignerAiry} results in identity with the Airy transform. Explicitly, we have $\alpha=\left(\frac{2\gamma}{\hbar}\right)^\frac13$ for $n=3$ and $\alpha=\left(\frac{2\cdot3q\gamma}{\hbar}\right)^\frac13$ for $n=4$. That is, if $\rho$ is a density operator and $W(q,p)$ its corresponding Wigner function then the transformation of the density operator $\rho\rightarrow U_n\rho U_n^\dagger$, $n=3,4$, corresponds to an Airy transform of $W(q,p)$ with respect to $p$ in the form
\begin{equation}
       W(q,p)\rightarrow\mathcal{A}_{\frac{\hbar}{2}\alpha}[W]\left(\mathcal{S}_n(p)\right)\,,
\end{equation}
where we have written the Airy transform such that standard transform pairs can be immediately used~\cite{vallee_airy_2010}. This shows that the action of cubic and quartic phase gates can be explicitly calculated in phase space. In the case that an impure $\rho$ is explicitly decomposed into a convex mixture of states, the linearity of the Wigner function still allows for the direct calculation of the effect of the cubic and quartic phase gates. For comparison with direct integration of the Wigner function see Appendix~II. Note that for $n>4$ the nonlinear phase gate exits the Airy form. This can be seen by considering the case $n=5$. The new exponential terms in the Wigner integral (c.f. Eq.~\ref{cubphas}) can then be expressed as
\begin{equation}
    e^{\frac{2i}{\hbar}\frac{\gamma t^5}{5}}e^{\frac{2i}{\hbar}2\gamma q^2t^3}e^{\frac{2it}{\hbar}\mathcal{S}_4(p)}\,.
\end{equation}
The rightmost term is linear in the integration variable $t$ and corresponds to a transformation of the phase space variables. The middle term, cubic in $t$, is an extra phase that prevents interpretation as a Wigner function, but that may be interpreted as an Airy transform, as for the cubic and quartic phase gates. However the leftmost term contains $t^5$ which prevents any interpretation as an Airy transform. Despite this difficulty, many higher order operations can be decomposed into a universal gate set containing either cubic or quartic gates, each member of which corresponds either to the Airy transforms we have described or the well-known symplectic transformations associated with Gaussian operations~\cite{kalajdzievski_exact_2021}. Therefore the Wigner functions resulting from higher order operations can still be obtained, given that both the Airy transform exists and the decomposition into the cubic or quartic gates can be found.

The Wigner functions after the application of the nonlinear phase gates for some specific initial states now follow directly from standard Airy transforms~\cite{vallee_airy_2010}. Firstly, if the initial state is the ideal momentum eigenstate, then we have the Wigner function $W(q,p)=\delta(p)$. Using $\mathcal{A}_\alpha[\delta](x)=\text{Ai}\left(x;\alpha\right)$ the Wigner function after the nonlinear phase gate is
\begin{equation}
    W(q,p)=\text{Ai}\left(\mathcal{S}_n(p);\alpha\right)\,,
\end{equation}
which compares favourably with a direct calculation when $\alpha$ is selected as detailed above. The Airy transform of the Gaussian function $f(x)=\frac{1}{\sqrt{2\pi}\sigma}e^{-\frac{(x-\mu)^2}{2\sigma^2}}$ is 
\begin{equation}\label{AiryGauss}
    \mathcal{A}_\alpha[f](x)=\frac{1}{|\alpha|}e^{\frac{2\sigma^2}{4\alpha^2}\left(\frac{x+\mu}{\alpha}+\frac{\sigma^4}{6\alpha^4}\right)}\text{Ai}\left(\frac{x+\mu}{\alpha}+\frac14\left(\frac{\sigma}{\alpha}\right)^4\right)\,.
\end{equation}
That is, the Wigner function of any initial Gaussian state must be transformed by the cubic or quartic gates into an Airy function of the phase space variables. A full decomposition of any pure or mixed Gaussian state in terms of the mean values and covariance matrix elements is given in the methods section below (see also~\cite{riera-campeny_wigner_2024}).

{\it Observation}: It follows directly from the analytical form of the post-gate Gaussian states that the cubic or quartic phase gates produce negativity in the Wigner function regardless of the impurity of the initial Gaussian state, even if the negative volume is vanishingly small. This can be seen from the fact that the factors multiplying the Airy function are always positive, whereas the Airy function itself must always be negative at some point. 
This analytical result is striking as such extraordinary robustness is difficult to see numerically because negative values become close to zero quickly (see Appendix~I), exemplifying the power of the analytical method. Note the contrast with the semiclassical squeezing effect of the Gaussian quadratic phase gate, which vanishes for some thermal distribution. 

We now turn to the nonlinear phase gate acting on a multimode state. Suppose that we have an $N$-mode state $\varrho$ and we wish to evaluate the action of $U_n$ on mode $i$. The Wigner function is expressed as
\begin{align}
\begin{split}
    &W(\mathbf{q},\mathbf{p})=\\
    &\int_{\mathbb{R}^N}\frac{e^{\frac{2i}{\hbar}\mathbf{p}\cdot\mathbf{t}}}{(\pi\hbar)^N}e^{-i\frac{\gamma}{\hbar n}((q_i-t_i)^n-(q_i+t_i)^n)}\braket{\mathbf{q}-\mathbf{t}|\varrho|\mathbf{q}+\mathbf{t}}d\mathbf{t}
\end{split}
\end{align}
where we have upgraded the phase space quantities to vectors in $\mathbb{R}^N$. For $\mathbf{v}\in\mathbb{R}^N$ let $\mathbf{v}^i\equiv\mathbf{v}\setminus\{v_i\}\in\mathbb{R}^{N-1}$. Then we may write
\begin{align}
\begin{split}
    &W(\mathbf{q},\mathbf{p})=\frac{1}{\pi\hbar}\int e^{\frac{2i}{\hbar}p_it_i}e^{-i\frac{\gamma}{\hbar n}((q_i-t_i)^n-(q_i+t_i)^n)}\times\\
    &\left(\frac{1}{(\pi\hbar)^{N-1}}\int_{\mathbb{R}^{N-1}}e^{\frac{2i}{\hbar}\mathbf{p}^i\cdot\mathbf{t}^i}\braket{\mathbf{q}-\mathbf{t}|\varrho|\mathbf{q}+\mathbf{t}}d\mathbf{t}^i\right)dt_i
\end{split}
\\ &=\mathcal{A}_{\alpha}[W](\mathcal{S}_n(p_i))\,,
\end{align}
where $\alpha$ must be chosen appropriately. That is, the action of the nonlinear phase gate corresponds to an Airy transform of the multimode Wigner function with respect to the target momentum variable. The only explicit example we know of is the cubic-phase entangled (CPE) state~\cite{mcconnell_unconditional_2023}, which we recalculate in Appendix~IV using this method. Decomposition of the quartic phase gate $U_4$ (involving an ancilla mode, and therefore requiring multimode analysis) and the continuous variable Toffoli gate $e^{i\frac{\gamma}{\hbar}\hat q_1\hat q_2\hat q_3}$ may be given in terms of a universal gate set involving the cubic phase gate, as well as many others~\cite{kalajdzievski_exact_2021}.

The universal gate set for continuous variable quantum computation can in principle now be implemented entirely in phase space, with the linear phase gates and the Fourier transform corresponding to linear symplectic transformations of the phase space variables and the cubic or quartic phase gates corresponding to Airy transforms of the Wigner function with respect to the nonlinear symplectic transformation of the phase space variables. We also note the connection that linear transformations can be implemented via convolution of the Wigner function with a Gaussian function, whereas for cubic and quartic phase gates the correct transformation is achieved via convolution with an Airy function.

\subsection{Application: Quartic-Bounded Cubic Gates}

\begin{figure}[h]
    \centering
    \includegraphics[width=\columnwidth]{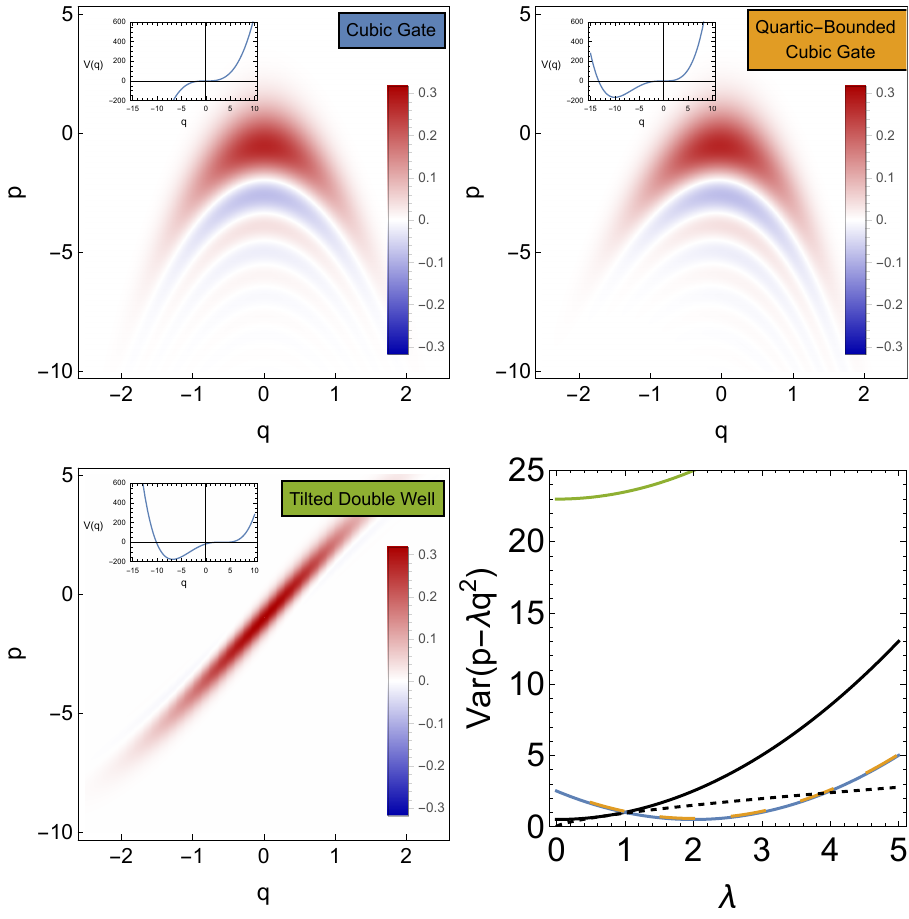}
    \caption{Application of Airy transforms to analyse physical bounded nonlinear phase gates. The upper row shows the cubic phase gate $U_3$ with $\gamma_3=2$, and the quartic-bounded cubic phase gate $U_{3,4}$ with $\gamma_3=2$ and $\gamma_4=0.2$. The bottom row shows the TDW gate generated by the unitary operator $U_\text{TDW}=\exp[-\frac{i}{\hbar}(-18+15\hat{q}-\frac{7}{2}\hat{q}^2+\frac{0.2}{4}\hat{q}^4)]$ with the same $\gamma_4$, approximating the quartic bounded cubic potential. The effect of lower bounding the cubic gate is to limit the dynamics for negative position and momentum. The TDW is a poor substitute for the cubic phase gate at the level of phase space representation, also reflected in the nonlinear squeezing. All gates take the harmonic oscillator ground state as the initial state with $\hbar=1$ and the insets are the equivalent cubic, quartic and tilted double well potentials $V(q)$ forming the gates. For the nonlinear squeezing the black dashed line is the threshold for quantum non-Gaussianity, solid black is the harmonic oscillator ground state, blue (orange) is the (quartic-bounded) cubic phase state and green is the TDW state.}
    \label{CvQ}
\end{figure}

Since the cubic and quartic phase gates commute it is possible to repeat this calculation for their combination, and find the transformation in phase space corresponding to the physical, lower-bounded, unitary transformation $U_{3,4}=e^{-\frac{i}{\hbar}\left(\frac{\gamma_3}{3\hbar}\hat q^3+\frac{\gamma_4}{4\hbar}\hat q^4\right)}$, representing a realistic unitary cubic gate. In this case, we find that the Wigner function, in the form congruent with Eqs.~\ref{WignerAiry}, may be written
\begin{multline}
    W(q,p)\xrightarrow{U_{34}}\frac{1}{\pi\hbar|\alpha|}\int~e^{\frac{2i\mathcal{S}_{3,4}(p)z}{\alpha\hbar}}e^{\frac{2i}{\alpha^3\hbar}\left(\frac{\gamma_3}{3}+\gamma_4q\right)z^3}\\\times\braket{q+\frac{z}{\alpha}|\rho|q-\frac{z}{\alpha}}dz\,,
\end{multline}
where we identify $\mathcal{S}_{3,4}(p)=p+\gamma_3 q^2+\gamma_4 q^3$ as the transformation of $\hat{p}$ associated with $U_{3,4}$. This is indeed another Airy transform with $\alpha=\left[\left(\frac{3\cdot2}{\hbar}\right)\left(\frac{\gamma_3}{3}+\gamma_4q\right)\right]^\frac13$. For $\gamma_4\ll \gamma_3$ we obtain a quartic-bounded cubic phase gate, which represents a transformation in a physical lower-bounded potential~\cite{felicetti_universal_2020,minganti_non-gaussian_2023}. That is, the transformation for a quartic-bounded cubic phase gate, including any lower order imperfections, for any input Wigner function can be obtained.

In Fig.~\ref{CvQ} we examine the effect of this quartic bounding on the creation of the cubic phase state. We show that the parabolic shape induced by the cubic semi-classical dynamics, as well as the quantum interference and negative regions are preserved, while the diverging negative momentum and position due to the cubic nonlinearity are suppressed by the quartic one. In contrast, a tilted double well (TDW) gate (with the same $\gamma_4$) designed to mimic the quartic-bounded cubic gate fails to generate anything like these features. This can be observed directly by the nonlinear squeezing~\cite{moore_quantum_2019}, which shows that the quartic-bounded cubic gate is a good approximation to the cubic gate, while the TDW is far above the quantum non-Gaussianity threshold.

While the major features of the cubic phase state are preserved when using the quartic-bounded cubic gate, errors due to the bounding can accumulate. A fixed unbounded cubic phase gate can be reversed (using our universal gate set) by applying a second gate sandwiched between a double Fourier transform, effectively generating the inverse cubic phase gate by changing the sign of $\gamma$. However the bounding quartic term does not change sign, and thus accumulates. The most natural way to solve this is to engineer the more difficult inverted quartic potential which itself must be bounded by higher order potentials. These will themselves accumulate, defining the principal limit of such gates and simulations in phase space. 

\begin{figure}[h]
    \centering
    \includegraphics[width=\columnwidth]{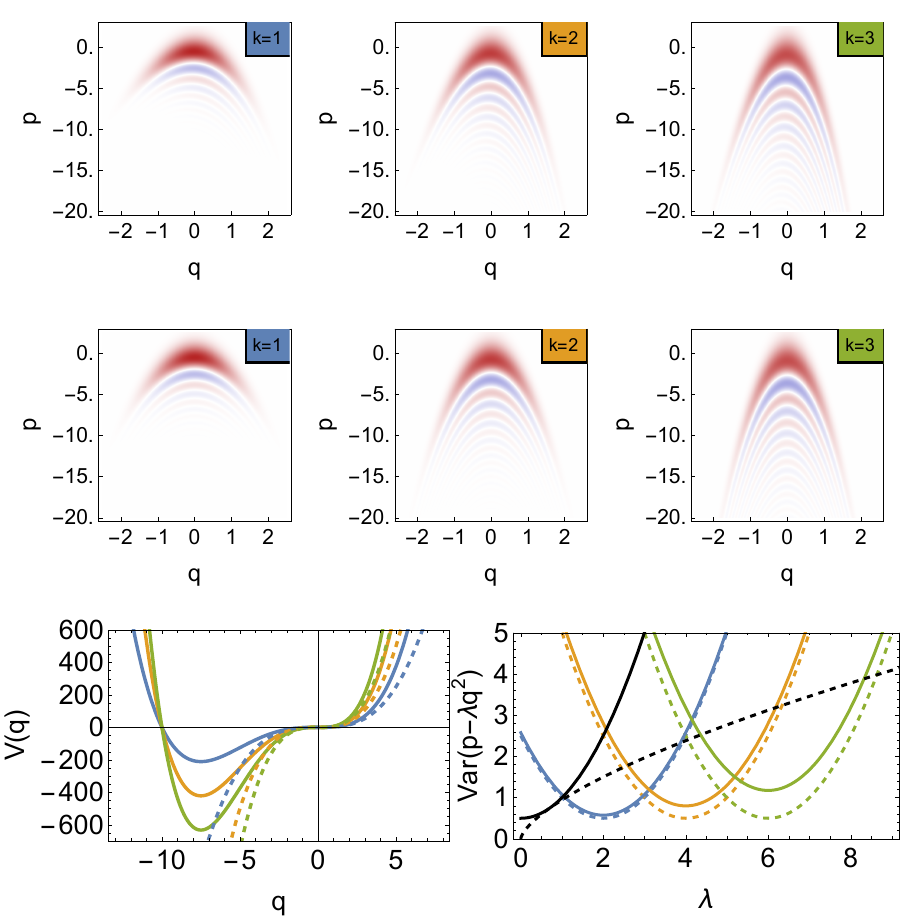}
    \caption{Application of Airy transform to analyse nonlinearity accumulation with physical bounded gates. Action of the quartic bound cubic gate $(U_{3,4})^k$ (top) for $k=1,2,3$ left to right. Iteration of the gate increases the cubic effects, as seen by the re-emergence of the suppressed negative position and momentum region. This occurs even though at all stages the system is bounded from below by the quartic gate. The pure cubic phase states (bottom) have diverging momentum for both positive and negative momentum symmetrically. Initial states and parameters are as in Fig.~\ref{CvQ}. The effective potentials corresponding to the applied gates are shown below the Wigner functions (left) where quartic-bounded cubic potentials are solid, cubic potentials are dashed. Increasing $k$ (blue, orange, green) leads to quartic-bounded cubic potentials that more closely approximate the cubic potential around the inflection point. However the increasing significance of the quartic term more strongly attenuates the nonlinear squeezing in absolute value in comparison with a cubic phase state (bottom right).}
    \label{Accum}
\end{figure}

An illustrative example is provided by considering a gate decomposition with the cubic phase gate realistically bounded by a weak quartic potential. One of the simplest such gate decompositions~\cite{kalajdzievski_exact_2021} is the multimode gate
\begin{equation}
    e^{i\frac{\gamma}{3}(q_j+q_k)^3}=F_je^{2iq_jq_k}F_j^\dagger e^{i\frac{\gamma_3}{3} q_j^3}e^{-2iq_jq_k}\,.
\end{equation}
The bounding quartic term in the Airy transform introduces extra terms that depend on $q$ through both the parameter $\alpha$ and the nonlinear transformation of $p$. Then, subsequent linear gates act nontrivially on these extra terms. We will use ideal states to suppress unwieldy calculations and demonstrate the principle. Concretely, applying this gate to a pair of zero-mean ideal momentum eigenstates produces the Wigner function
\begin{multline}
    W_\text{C}=\text{Ai}\left(p_j-\gamma_3(q_j+2q_k)^2+2q_k;\alpha\right)\times\\
    \delta\left(p_k+2(2q_k+q_j-p_j)\right)\,.
\end{multline}
In evaluating the effect of the quartic bounded cubic phase gate we write $\alpha\equiv\alpha(q)$ in order keep track of the nontrivial $q$ variable added by the quartic term. This then produces the Wigner function
\begin{multline}
    W_\text{QBC}=\text{Ai}\left(p_j-\gamma_3Q_{jk}+\gamma_4Q_{jk}^3+2q_k;\alpha(Q_{jk})\right)\\\times\delta\left(p_k+2(Q_{jk}-p_j)\right)\,,
\end{multline}
where $Q_{jk}=q_j+2q_k$.

Despite this, repeated application of the quartic bounded cubic gate results in Wigner functions that are dominated by cubic rather than quartic effects. The gate $(U_{3,4})^k$ simply accumulates cubic and quartic terms, so that $\gamma_{3,4}\rightarrow k\gamma_{3,4}$. Fig.~\ref{Accum} shows the progression from $k=1$ to $k=3$. Even though at each step the cubic potential is bounded from below by the quartic potential, the cubic effects are enhanced by repeated application with diverging negative momentum and position reasserting themselves as $k$ increases.

\subsection{Unbounded Dynamics}\label{Unbound}

\begin{figure}[h]
    \centering
    \includegraphics[width=\columnwidth]{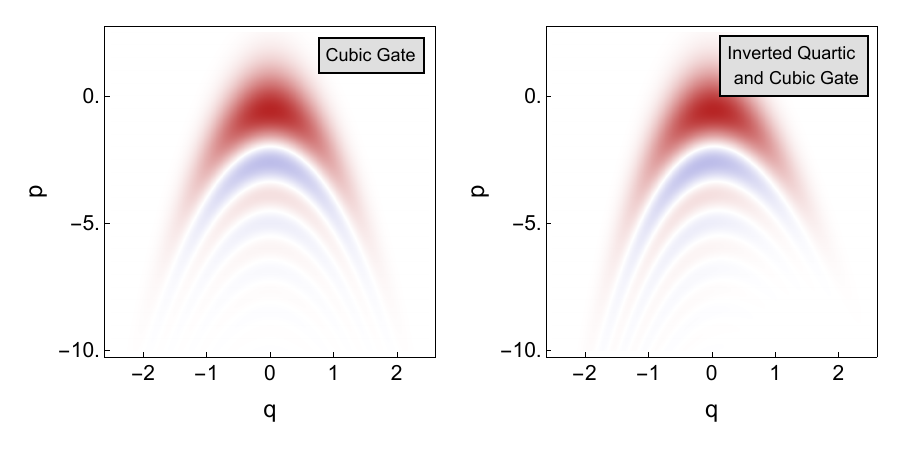}
    \caption{Application of Airy transform to analyse a realistic cubic nonlinearity softening. The bare cubic gate compared with the cubic gate softened for $q>0$ by a weak inverted quartic nonlinearity. The diverging negative position and momentum due to the cubic gate are present in both examples. The effect of a weak inverted quartic gate is to suppress the positive position and negative momentum. The divergence is then faster in the region where both cubic and quartic potentials go to negative infinity together. Initial states and parameters are as in Fig.~1 of the main text.}
    \label{Unbound}
\end{figure}

We note that our methodology can also be used to probe fully unbounded nonlinear dynamics such as the inverted quartic potential in the large mass limit. In the case where $\gamma_3=2$ (as before) and $\gamma_4=-0.2$ the divergence into negative momentum and position from the cubic potential is no longer constrained by the hardening wall of the quartic potential. Instead, since this region of the potential now softens faster than the region including the hardening wall of the cubic potential, this divergence returns and the divergence on hard cubic side is suppressed. The corresponding Wigner functions are shown in Fig.~\ref{Unbound}.

\section{Methods}
Our application to quartic bounded cubic phase gates and our observation of the resilience of negativity to initial thermal noise rely on performing the required Airy transform on Gaussian states. Here we fully outline this method.

\subsection{Wigner Function of the Cubic and Quartic Phase States}\label{CQstates}

For ease of reference we will refer to the set of states generated by the nonlinear phase gates $U_3$, $U_4$ or $U_{3,4}$ acting on an arbitrary Gaussian state $\rho_G$ as the cubic, quartic or cubic and quartic phase states. Additionally, it will be useful to introduce the Fock basis through the annihilation operator $a$ where $\hat{q}=\sqrt{\frac{\hbar}{2}}(a+a^\dagger)$ and $\hat{p}=\sqrt{\frac{i\hbar}{2}}(a^\dagger-a)$. This may be understood as establishing a reference oscillator with mass and frequency set to unity. Formally, these nonlinear phase states may be expressed as the sets of density operators
\begin{equation}
    \mathcal{G}_A=\{U_A\rho_GU_A^\dagger|\gamma\in\mathbb{R}~\land~\rho_G=D(\beta)S(z)\nu S(z)^\dagger D(\beta)^\dagger\}\,,
\end{equation}
where $A=3,4$ or the pair $(3,4)$, $D(\beta)=e^{\beta a^\dagger-\beta^*a}$ is the displacement operator where $\beta=\frac{\bar{q}+i\bar{p}}{\sqrt{2}\hbar}$ contains the mean values of $\hat q$ and $\hat p$, $S(z)=e^{\frac12\left(z(a^\dagger)^2-z^*a^2\right)}$ is the squeezing operator with $z=re^{i\theta}$, and $\nu=\frac{1}{1+\bar{n}}\sum_k\left(\frac{\bar{n}}{1+\bar{n}}\right)^k\ket{k}\bra{k}$ is a Gaussian thermal state, characterised by a mean occupation $\bar{n}$~\cite{ferraro_gaussian_2005}.

The Wigner function of $\rho\in\mathcal{G}_A$ is the Airy transform of the Gaussian state $\rho_G$
\begin{equation}
    W[\rho](q,p)=\mathcal{A}_{\frac{\hbar}{2}\alpha}[W_G](\mathcal{S}_A(p))\,.
\end{equation}
From here, one may take the Wigner function of the thermal state $W_\text{th}(q,p)=\frac{e^{-\frac{(q^2+p^2)}{\hbar(1+2\bar n)}}}{(1+2\bar{n})\pi\hbar}$, perform the relevant symplectic transformations associated with displacement and squeezing, then perform the relevant Airy transform, bearing in mind the scaling and translation rules which state that state that if $\phi_\alpha(x)$ is the Airy transform of $f(x)$ then $\phi_{\alpha k}(kx)$ is the Airy transform of $f(kx)$ and $\phi_\alpha(x+s)$ is the Airy transform of $f(x+s)$. 

To rapidly gain access to a general expression for the phase states, it is useful to note that since $W_G$ is Gaussian it can be expressed in the matrix form
\begin{equation}
    W_G(\mathbf{x})=\frac{1}{2\pi\sqrt{\text{Det}(\Sigma)}}e^{-\frac12(\mathbf{x}-\mu)^\top\Sigma^{-1}(\mathbf{x}-\mu})\,,
\end{equation}
where the covariance matrix $\Sigma$ and the mean values $\mu$ take the generic forms
\begin{align}
    \Sigma&=\begin{pmatrix} \sigma_q & \sigma_{qp}\\
    \sigma_{qp} & \sigma_p\end{pmatrix}\\
    \mu&=\begin{pmatrix}
        \bar{x} \\ 
        \bar{p}
    \end{pmatrix}\,.
\end{align}
$W_G$ can be expanded in terms of these generic forms and the terms depending on $p$ factored, so that we have
\begin{multline}
    W_G(q,p)=\frac{e^{-\frac{\bar{x}^2\sigma_p+\bar{p}^2\sigma_q-2\bar{x}\bar{p}\sigma_{qp}}{2\text{Det}(\Sigma)}}}{2\pi\sqrt{\text{Det}(\Sigma)}}e^{-\frac{\sigma_pq^2+2(\bar{p}\sigma_{qp}-\bar{x}\sigma_pq)}{2\text{Det}(\Sigma)}}\\\times e^{-\frac{\sigma_qp^2+2(\bar{x}\sigma_{qp}-(\bar{p}\sigma_q+\sigma_{qp}q))p}{2\text{Det}(\Sigma)}}\,.
\end{multline}
In order to perform the Airy transform with respect to $p$, it is useful to complete the square for $p$, resulting in the expression
\begin{multline}
    \left(\sqrt{\frac{\sigma_q}{2\text{Det}(\Sigma)}}p-\frac{(\bar{x}\sigma_{qp}-(\bar{p}\sigma_q+\sigma_{qp}q))}{\sqrt{2\sigma_q\text{Det}(\Sigma)}}\right)^2\\-\frac{\left(\bar{x}\sigma_{qp}-(\bar{p}\sigma_q+\sigma_{qp}q)\right)^2}{2\sigma_q\text{Det}(\Sigma)}
\end{multline}

If $f(x)=\frac{e^{-x^2}}{\sqrt{\pi}}$ is the normalised Gaussian function then
\begin{equation}
    \mathcal{A}_\alpha[f](x)=\frac{1}{|\alpha|}e^{\frac{1}{4\alpha^3}\left(x+\frac{1}{24\alpha^3}\right)}\text{Ai}\left(\frac{x}{\alpha}+\frac{1}{16\alpha^4}\right)\,.
\end{equation}
The scaling and translation rules for Airy transforms state that if $\phi_\alpha(x)$ is the Airy transform of $f(x)$ then $\phi_{\alpha k}(kx)$ is the Airy transform of $f(kx)$ and $\phi_\alpha(x+s)$ is the Airy transform of $f(x+s)$, and thus combined $\phi_{\alpha k}(kx+s)$ is the Airy transform of $f(kx+s)$. Therefore selecting
\begin{align}
    k&=\sqrt{\frac{\sigma_q}{2\text{Det}(\Sigma)}}\\
    s&=-\frac{(\bar{x}\sigma_{qp}-(\bar{p}\sigma_q+\sigma_{qp}q))}{\sqrt{2\sigma_q\text{Det}(\Sigma)}}\,,
\end{align}
and performing the Airy transform with respect to $\mathcal{S}_A(p)$ retrieves the Wigner function of the nonlinear phase states.
The full expression for the Wigner function of an arbitrary Gaussian state following the nonlinear phase gate is
\begin{widetext}
    \begin{equation}\label{GaussWig}
        W_{\mathcal{G}_A}(q,p)=\frac{e^{-\frac{(q-\bar{x})^2}{2\sigma_q}}}{2\sqrt{\pi\text{Det}(\Sigma)}}\frac{1}{|\alpha^\prime|}e^{\frac{1}{4\alpha^\prime{}^3}\left(k\mathcal{S}_A(p)+s+\frac{1}{24\alpha^\prime{}^3}\right)}\text{Ai}\left(\frac{k\mathcal{S}_A(p)+s}{\alpha^\prime}+\frac{1}{16\alpha^\prime{}^4}\right)\,,
    \end{equation}
\end{widetext}
where $\alpha^\prime=\frac{k\hbar}{2}\alpha$. Appropriate selections of $A$ and $\alpha$ recover the various nonlinear phase states.

\section{Discussion}

We have presented a general method for evaluating the effect of nonlinear phase gates in phase space. We add that the effect of the cubic phase gate on the momentum probability distribution can be directly evaluated without dealing with the rather troublesome marginal integrals of the Wigner function, and some details of this are provided in Appendix~III. We intend to explore this connection in further work. The method focuses on phase gates built out of position operators, however as the Wigner function can be written in terms of momentum eigenstates, phase gates built out of momentum operators can also be accommodated by simply switching to this picture.

Some elementary extensions of this method may become possible in the future. As noted, the higher order phase gates ($n>4$) result in the integrand of the Wigner function possessing an exponential of polynomials of order greater than $3$, which no longer conforms to the Airy transform structure. Higher order generalisations of Airy functions do exist but do not deal with the retention of the lower order terms in the polynomial~\cite{durugo_higher-order_2014}, and anyway would require a theory of generalised Airy transforms to be constructed. For multimode extensions similar challenges arise. A true multimode extension of the cubic or quartic phase gates involves a nondegenerate cubic or quartic interaction among multiple modes. For cubic gates the two possibilities are represented by the operators $U_{\mathcal{C}2}=e^{i\gamma q_1q_2^2}$ and $U_{\mathcal{C}3}=e^{i\gamma q_1q_2q_3}$, with the latter being the generator of the lowest order continuous variable quantum hypergraph states~\cite{moore_quantum_2019} or the continuous variable Toffoli gate. The Wigner integrals involved in such calculations contain inhomogeneous forms of order 3. Such non-Gaussian integrals are notoriously difficult to solve, and yet progress has been made even in recent years for non-Gaussian integrals involving homogeneous forms~\cite{morozov_introduction_2009,shakirov_nonperturbative_2010}.  Another major roadblock using these methods is the non-commutativity of the $\hat q$ and $\hat p$ operators, with one of the most important applications being nonlinear motion. The introduction of such noncommutative operators even at the level of phase rotation can bring significant complexity to the Airy transform, particular after gate sequences involving multiple nonlinear phase gates.

Recent experimental achievements demonstrate the importance of the theoretical advance presented here. Cubic phase states have been produced in optical~\cite{kudra_robust_2021} and superconducting circuit~\cite{eriksson_universal_2023} settings. Alongside these achievements much effort has gone into theoretical proposals for the cubic phase states~\cite{bartlett_universal_2002,park_deterministic_2018,yanagimoto_engineering_2020,zheng_gaussian_2021,sakguchi_nonlinear_2021,houhou_unconditional_2022} and detailed theoretical studies assess the properties and suitability of cubic phase states for various applications~\cite{konno_nonlinear_2021,hastrup_unsuitability_2021,kalajdzievski_exact_2021,budinger_all-optical_2022,neumeier_fast_2022,riera-campeny_wigner_2024}. Clarifying the Wigner function for such states will help explore such properties and open paths to understand the unique forms of quantum interference they generate. Similarly, the quartic potential is an important and paradigmatic example of a nonlinear bounded potential, often appearing as a double well potential~\cite{wu_nonequilibrium_2022,roda-llordes_macroscopic_2023}, and can be used for quantum information tasks~\cite{weiss_quantum_2019}. Remarkably, the aforementioned superconducting microwave circuit experiment~\cite{eriksson_universal_2023} producing cubic phase states also allows for the simultaneous presence of trilinear and Kerr-like nonlinearities, underlining the importance of understanding the simultaneous presence of cubic and quartic nonlinearities. While we have focused on nonlinear gates acting on Gaussian states, our method applies equally well to Fock states or finite superpositions thereof. Hence there is also an opportunity to investigate the interaction of two opposing forms of nonlinearity in bosonic systems.

Note: During the last stage of manuscript preparation an independent manuscript~\cite{riera-campeny_wigner_2024} addressed a different problem of position delocalization in open mechanical dynamics that complements our analytical results on cubic and quartic unitary gates.
 
\begin{acknowledgments}
D.M. acknowledges the project 23-06224S of Czech Science Foundation, the European Union's HORIZON Research and Innovation Actions under Grant Agreement no. 101080173 (CLUSTEC) and the project
CZ.02.01.01/00/22 008/0004649 (QUEENTEC) of EU and MEYS Czech Republic. R.F. acknowledges the project 21-13265X of the Czech Science Foundation. We have further been supported by the European Union's 2020 research and innovation programme (CSA - Coordination and support action, H2020-WIDESPREAD-2020-5) under grant agreement No. 951737 (NONGAUSS).
\end{acknowledgments}

\section{Data Availability}
Data or code used in this study available upon reasonable request.

\section{Author Contributions}
D. M. developed the theory methods connecting nonlinear phase gates and integral transforms. R. F. conceived the applications of the theory methods, quantum non-Gaussian aspects, and their analysis. Both authors contributed to the discussions and writing of the manuscript.

\section{Competing Interests}
The authors declare no competing interests.


\appendix

\onecolumngrid

\section{Robustness of Wigner Negativity for Thermal States}\label{SurviveNeg}

The full expression for the Wigner function of the nonlinear phase states (see main text, Methods) allows us to directly consider the effect of initial Gaussian thermal noise on the negativity in the Wigner function induced by the nonlinear phase states. It follows directly from the analytical form of the $W_{\mathcal{G}_A}$ that the negativity must survive arbitrarily large initial thermal noise $\bar{n}$. The Wigner function is composed of a product of positive functions with an Airy function. Since the Airy function is always negative at at least one point, it follows that the Wigner function is always negative at at least one point. 

\begin{figure}
    \centering
    \includegraphics[width=\columnwidth]{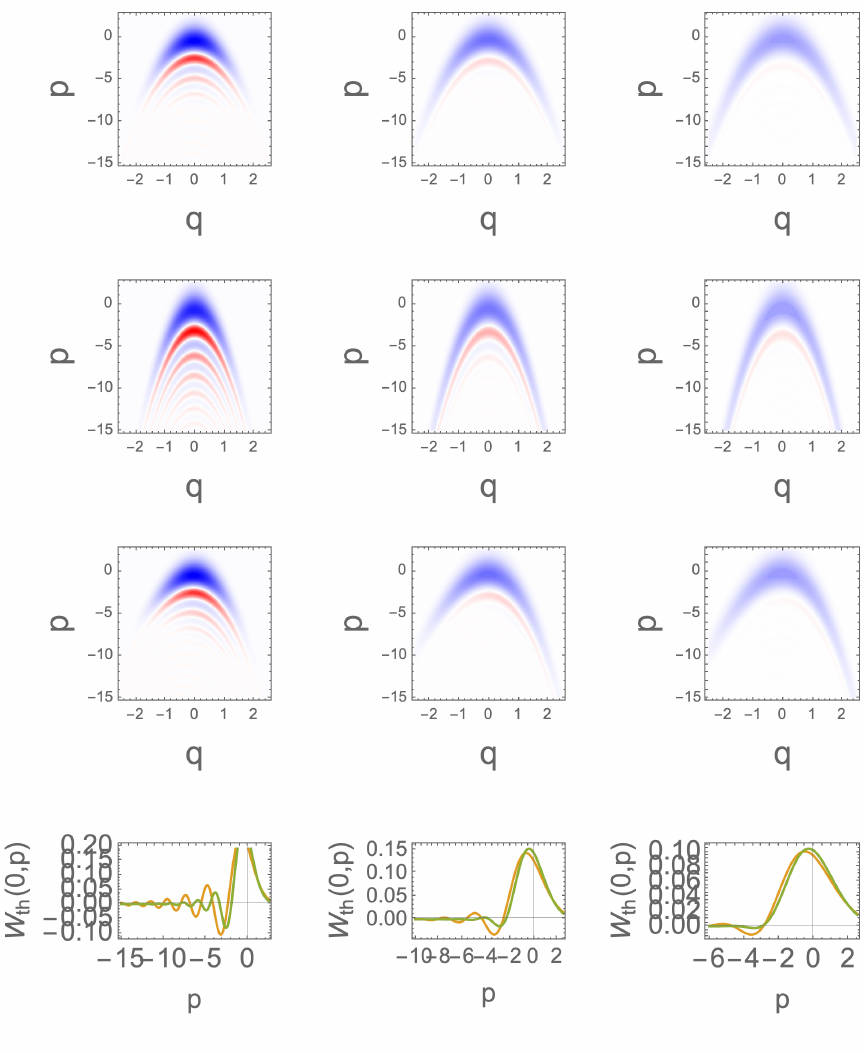}
    \caption{The Wigner functions for the cubic phase gate acting on various thermal states. Horizontally we set $\bar n=0, 0.5, 1$ and verically $\gamma_3=2$ (top), $\gamma_3=4$ (middle) and the quartic-bounded cubic gate (bottom) with $\gamma_3=2$ and $\gamma_4=0.2$. The nonclassical fluctuations are suppressed by the thermal noise, while the state spreads out in phase space, as seen by the extension of the parabolic `arms'. Below these we show the Wigner cuts at $q=0$ with blue, yellow and green following the vertical ordering.}
    \label{Thermals}
\end{figure}

To illustrate this more concretely consider that the cubic phase gate acting on a thermal state (see also Fig.~\ref{Thermals}) results in the Wigner function
\begin{widetext}
\begin{equation}
    W_\text{th}(q,p)=\frac{2^\frac23e^{\frac{4\bar{n}(1+\bar{n})q^2}{1+2\bar{n}}+\frac{(1+2\bar{n})^3+6(1+2\bar{n})\gamma p}{6\gamma^2}}\text{Ai}\left(\frac{1+4\bar{n}(1+\bar{n})+4\gamma(p+\gamma q^2)}{(2\gamma)^\frac43}\right)}{\sqrt{\pi(1+\bar{n})}|\gamma^\frac13|}\,.
\end{equation} 
\end{widetext} 
Due to the parabolic symmetry of this function we can take a cut of the Wigner function at $q=0$, which shows the negative ripples along the $p$ axis and has the form
\begin{widetext}
\begin{equation}
    W_\text{th}(0,p)=\frac{2^\frac23e^{\frac{(1+2\bar{n})^3+6(1+2\bar{n})\gamma p}{6\gamma^2}}\text{Ai}\left(\frac{1+4\bar{n}(1+\bar{n})+4\gamma p}{(2\gamma)^\frac43}\right)}{\sqrt{\pi(1+\bar{n})}|\gamma^\frac13|}\,.
\end{equation} 
\end{widetext}
For $p<0$, the Airy function always provides oscillations between positive and negative values for any finite $\bar n$ (see Fig.~\ref{Thermals}, bottom row). However the exponential term involving $p$ is a decaying function which reduces the amplitude of the oscillations, with the reduction proportional to $\bar n$. The Airy oscillations in the negative regions are thus suppressed to zero {\it from below} and reach zero in the limit $\bar n\rightarrow\infty$. Similar arguments can be applied to the results of other nonlinear phase gates. 

\section{Numerical Integration as compared with Airy Transforms}\label{NumInt}

Interpreting the cubic and quartic phase gates as Airy transforms potentially provides some computational advantage, in addition to the broad analytical results mentioned in the main text. Here we illustrate briefly in what form this advantage might take by comparing the direct numerical integration of the Wigner function with the analytical result. In fact, the simplest possible example comes from `cuts' or `slices' of the Wigner function and in this case the advantage of the analytical expressions is already visible.

\begin{figure}
    \centering
    \includegraphics[width=0.49\columnwidth]{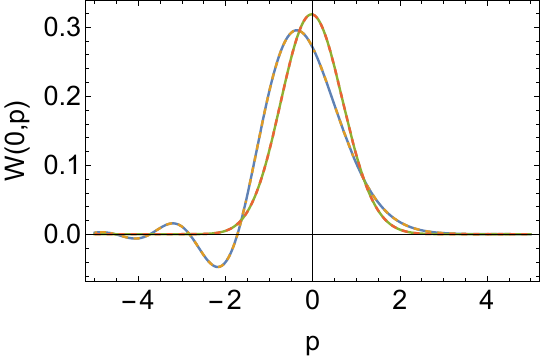}
    \includegraphics[width=0.49\columnwidth]{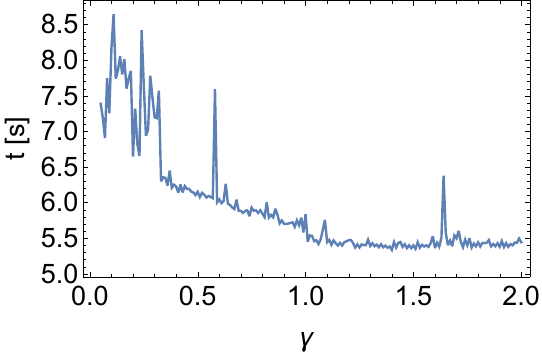}
    \caption{The Wigner cut $W(0,p)$ for the cubic phase state with $\gamma=0.05,1$. The accuracy with respect to the analytical calculation is very good, although some values for $\gamma=0.05$ were erroneously reported as complex-valued (not shown). Curves from numerical integration are dashed. The lower values of $\gamma$ appear to require more time to estimate numerically.}
    \label{WigCutCub}
\end{figure}

Consider the cubic phase state $U_3\ket{0}$, which has the Wigner function (with $\hbar=1$)
\begin{align}
    W(q,p)&=\frac{1}{\pi}\int e^{2ipt}e^{2i\frac{\gamma t^3}{3}}\frac{e^{-\frac12(q-t)^2}}{\pi^\frac14}\frac{e^{-\frac12(q+t)^2}}{\pi^\frac14}\\
    &=\frac{2^\frac23e^{\frac{1+6\gamma p}{6\gamma^2}}}{\sqrt{\pi}|\gamma|^\frac13}\text{Ai}\left(\frac{1+4\gamma\mathcal{S}_3(p)}{(2\gamma)^\frac43}\right)\,.
\end{align}
Then the Wigner cut $W(0,p)$ can be calculated both directly and by numerical integration. We do not aim for a detailed numerical analysis, or take advantage of specialised techniques relevant to the kind of integral being calculated. Rather, we straightforwardly compare the times required for the common mathematical software Mathematica using the command NIntegrate on its default settings to evaluate 101 evenly spaced points of the Wigner cut in the range $-5\le p\le5$. The time taken is estimated using AbsoluteTiming. Strikingly, the time is greater for low values of $\gamma$, as demonstrated in Fig.~\ref{WigCutCub}.

\begin{figure}
    \centering
    \includegraphics[width=0.49\columnwidth]{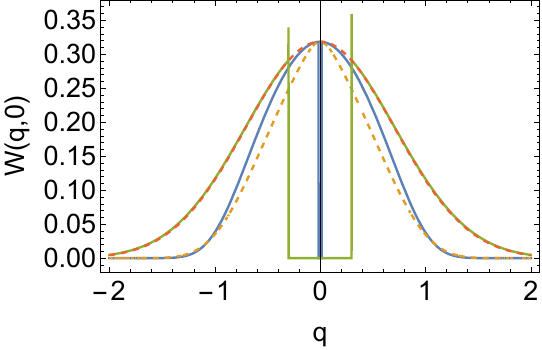}
    \includegraphics[width=0.49\columnwidth]{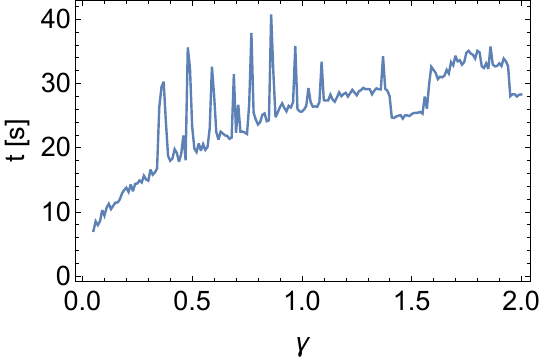}
    \caption{The Wigner cut $W(q,0)$ for the quartic phase state with $\gamma=0.05,1$. The accuracy with respect to the analytical calculation very good, although again some values for $\gamma=0.05$ were erroneously reported as complex-valued (not shown). Curves from numerical integration are dashed. The higher values of $\gamma$ appear to require more time to estimate numerically in contrast with the cubic state. In this case, the plotting software fails for the analytical expression as some numbers are too small to be represented numerically.}
    \label{WigCutQ}
\end{figure}

The quartic phase state $U_4\ket{0}$ has the Wigner function
\begin{align}
    W(q,p)&=\frac{1}{\pi}\int e^{2ipt}e^{2iq\gamma t^3}\frac{e^{-\frac12(q-t)^2}}{\pi^\frac14}\frac{e^{-\frac12(q+t)^2}}{\pi^\frac14}\\
    &=\frac{2^\frac23e^{\frac{1+6\gamma pq}{6(q\gamma)^2}}}{\sqrt{\pi}|q\gamma|^\frac13}\text{Ai}\left(\frac{1+4q\gamma\mathcal{S}_4(p)}{(2q\gamma)^\frac43}\right)\,.
\end{align}
Consider the Wigner cut $W(q,0)=\frac{2^\frac23e^{\frac{1}{6(q\gamma)^2}}}{\sqrt{\pi}|q\gamma|^\frac13}\text{Ai}\left(\frac{1+4\gamma^2q^4}{(2q\gamma)^\frac43}\right)$. In fact, this immediately poses numerical problems for Mathematica, due to the denominator when $q\rightarrow0$. While Mathematica evaluates $\lim_{q\rightarrow0}W(q,0)=\frac{1}{\pi}$, the plotting software fails to capture this (Fig.~\ref{WigCutQ}), due to some expressions involving numbers that are too small to be represented numerically. More precisely, the argument of the Airy function can be large even near the phase space origin, and the evaluation of the function at such values is a small number. The numerical integration also seems to take significantly longer than the cubic state, however it does seem to be able to avoid some of the problem of small numbers.

\begin{figure}
    \centering
    \includegraphics[width=0.49\columnwidth]{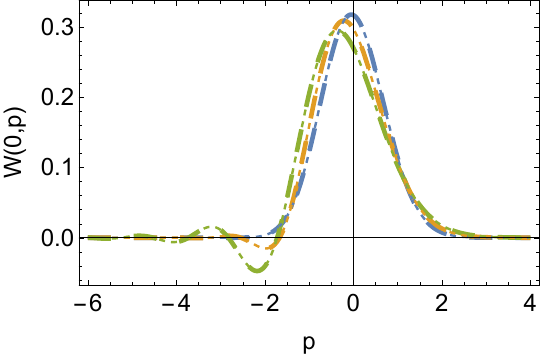}
    \includegraphics[width=0.49\columnwidth]{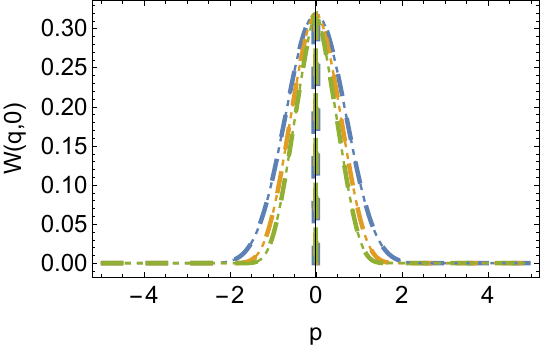}
    \caption{The Wigner cut $W(0,p)$ for the cubic phase state (left) and $W(q,0)$ for the quartic phase state with $\gamma=0.1$ (blue), $\gamma=0.5$ (yellow) and $\gamma=1$ (green). The thick dashed lines are from analytical expressions and the thin dashed lines are from numerical integration. The accuracy is high for these examples, although analytical expressions are much faster to plot, for example the Wigner cut plot for the cubic phase state using AbsoluteTiming takes approximately 0.03 seconds while the numerically integrated cut using NIntegrate takes approximately 30 seconds.}
    \label{Acc}
\end{figure}

We can also take note of the accuracy differences between the analytical calculations and numerical integration. In Fig.~\ref{Acc} we show Wigner cuts for the cubic and quartic phase states from the analytical calculation and numerical calculation. The accuracy of the numerical integration is high for these examples, but the time taken to produce the plots is much higher although certainly not prohibitive.

\section{Probability Distributions of the Canonical Variables} \label{ProbNonG}

In the Heisenberg picture it is easy to see that the variable $\hat{q}$ is invariant under the action of the cubic phase gate, with the obvious consequence that if the initial probability distribution of $\hat{q}$ is known then it remains known after the gate application. The same is not true of the distribution associated with $\hat{p}$ which becomes drastically non-Gaussian; straightforwardly, the calculation of the probability amplitude for $\hat{p}$ for the ideal cubic phase state involves the integral
\begin{align}
    \braket{p|e^{i\frac{\gamma}{3}\hat{q}^3}|0}_p&=\int\braket{p|x}e^{i\frac{\gamma}{3}x^3}\braket{x|0}_pdx\\
    &=\frac{1}{2\pi}\int e^{ixp}e^{i\frac{\gamma}{3}x^3}dx\\
    &=\frac{\gamma^{-\frac13}}{\pi}\int_0^\infty e^{i\gamma^{-\frac13}px}e^{ix^3}dx\\
    &=\gamma^{-\frac13}\text{Ai}(\gamma^{-\frac13}p)\,,
\end{align}
where we have rescaled the integration variable $x\rightarrow\gamma^{-\frac13}x$.

The cubic phase gate can be interpreted as applying an Airy transform to the momentum wavefunction. While the probability could also be calculated as the marginal of the associated Wigner distribution, the Airy transform method can bypass the potentially troublesome integral of an Airy function. 

For pure states, the momentum space wavefunction $\phi(p)=\frac{1}{\sqrt{2\pi\hbar}}\int e^{\frac{-ixp}{\hbar}}\psi(x)dx$ is just the Fourier transform of the position wavefunction $\psi(x)$. For an arbitrary initial state $\ket{\psi}$ we have that the effect of the cubic phase gate $U_3$ on the position wavefunction is to add a cubic phase, $\braket{x|U_3|\psi}=e^{-i\frac{\gamma x^3}{3\hbar}}\psi(x)$. Then the momentum wavefunction is
\begin{align}
    \braket{p|U_3|\psi}&=\mathcal{F}\{e^{-i\frac{\gamma x^3}{3\hbar}}\psi(x)\}\\
    &=\mathcal{F}\{e^{-i\frac{\gamma x^3}{3\hbar}}\}\ast\mathcal{F}\{\psi(x)\}\\
    &=\text{Ai}(p;\alpha)\ast\phi(p)\,,
\end{align}
where we used the convolution theorem and the Fourier pair for the Airy function. What remains is to specify $\alpha$. If we examine the momentum wavefunction in detail we find
\begin{align}
    \braket{p|U_3|\psi}&=\frac{1}{\sqrt{2\pi\hbar}}\int e^{-i\left(\frac{\gamma x^3}{3\hbar}+\frac{xp}{\hbar}\right)}\psi(x)dx\\
    &=\frac{1}{2\pi\hbar}\int e^{-i\left(\frac{\gamma x^3}{3\hbar}+\frac{xp}{\hbar}\right)}\int e^{i\frac{xt}{\hbar}}\phi(t)dxdt\\
    &=\frac{1}{2\pi\hbar}|\hbar\gamma^{-1}|^{\frac13}\int e^{i\left(\frac{x^3}{3}-\left(\hbar^2\gamma\right)^\frac13x(t-p)\right)}\phi(t)dxdt\,,
\end{align}
where we used the Fourier pairing between position and momentum wavefunctions and rescaled the integration variable $x=-\left(\hbar\gamma^{-1}\right)^\frac13$. It follows that we may identify $\alpha=-(\hbar^2\gamma)^{-\frac13}$. The effect of the cubic phase gate on the momentum wavefunction therefore is to convolve it with an Airy function, which is the definition of the Airy transform of $\phi(p)$. The probability distribution is then the modulus squared of this new wavefunction. Equivalently, the momentum probability distribution is a product of Airy transforms
\begin{align}
    \mathcal{P}(p)&=\braket{p|U_3|\psi}\braket{\psi|U_3^\dagger|p}\\
    &=\text{Ai}(p)\ast\phi(p)\cdot\text{Ai}(p)\ast\phi^*(p)\,.
\end{align}

In the case that the state is not pure there are no wavefunctions and instead the probability distribution is given by expectation values of the density matrix with the projectors of the canonical variables. That is, $\mathcal{P}(p)=\Tr\left(\ket{p}\bra{p}U_3\rho U_3^\dagger\right)=\braket{p|U_3\rho U_3^\dagger|p}$. One may decompose the density matrix as a convex sum of pure states $\rho=\sum_ip_i\ket{\psi_i}\bra{\psi_i}$. Then the probability distribution is just a convex sum of convolutions of the respective momentum wavefunctions with Airy functions
\begin{equation}
    \mathcal{P}(p)=\sum_ip_i\text{Ai}(p)\ast\phi_i(p)\cdot\text{Ai}(p)\ast\phi_i^*(p)\,.
\end{equation}

This method does not generalise to the quartic phase gate as there is no cancellation of the fourth order integration variable as in the Wigner function.

\section{Wigner Function of the CPE State}\label{CPE}

In Ref~\cite{mcconnell_unconditional_2023} the Wigner function of the CPE state is calculated explicitly. Here we show that we recover this result using the Airy transform formalism. The state is defined as
\begin{equation}
    \ket{z,\gamma,\theta}=U_3U_{BS}(\theta)\left[S(z)\otimes S^\dagger(z)\right]\ket{0}^{\otimes2}\,,
\end{equation}
where $U_{BS}(\theta)=e^{i\frac{\theta}{\hbar}(\hat{q}_1\hat{p}_2-\hat{q}_2\hat{p}_1)}$ is a beamsplitter operation and $S(z)$ is the squeezing operator (see Wigner Function of the Cubic and Quartic Phase States in the main text). Select $\theta=\frac{\pi\hbar}{4}$ and $z=\ln r>0$ to recover the state from the reference, although for simplicity here both squeezing operations are assumed to have equal strength. The Gaussian operators induce symplectic transformations on the phase space variables represented by the symplectic matrices:
\begin{align}
    S_{S(r)}&=\text{diag}\left(r,\frac{1}{r}\right)\\
    S_{U_{BS}(\frac{\pi}{4})}&=\frac{1}{\sqrt{2}}\begin{pmatrix}
        1 & 1 & 0 & 0\\
        1 & -1 & 0 & 0\\
        0 & 0 & 1 & 1\\
        0 & 0 & 1 & -1
    \end{pmatrix}\,.
\end{align}
Therefore the effect of the squeezing and beamsplitter operations can be calculated by substitution of these transformations into the Wigner function of the product of two harmonic oscillator ground states. The result is the entangled Gaussian two-mode squeezed state with Wigner function
\begin{equation}
    W_\text{TMSS}(\mathbf{q},\mathbf{p})=\frac{e^{-\frac{(q_1-q_2)^2+(p_1+p_2)^2+((p_1-p_2)^2+(q_1+q_2)^2)r^4}{2\hbar r^2}}}{(\pi\hbar)^2}\,.
\end{equation}
To apply the cubic phase gate, we need only perform the Airy transform with respect to the variable $p_1$. Much of the above expression factors out of the Airy transform, and we are left with
\begin{widetext}
\begin{equation}
    W_\text{CPE}(\mathbf{q},\mathbf{p})=\sqrt{\pi}\frac{e^{-\frac{p_2^2+(q_1-q_2)^2+(p_2^2+(q_1+q_2)^2)r^4}{2\hbar r^2}}}{(\pi\hbar)^2}e^{-\frac{(1-r^4)^2p_2^2}{2\hbar r^2(1+r^4)}}\mathcal{A}_{\frac{\hbar}{2}\left(\frac{2\gamma}{\hbar}\right)^\frac13}\left[\frac{e^{-\left(\sqrt{\frac{1+r^4}{2\hbar r^2}}p_1+\frac{p_2(1-r^4)}{\sqrt{2\hbar r^2(1+r^4)}}\right)^2}}{\sqrt{\pi}}\right]\left(\mathcal{S}_3(p_1)\right)\,.
\end{equation}
\end{widetext}

This is an Airy transform of a Gaussian function, where we have completed the square for $p_1$. The scaling and translation rules for Airy transforms state that if $\phi_\alpha(x)$ is the Airy transform of $f(x)$ then $\phi_{\alpha k}(kx)$ is the Airy transform of $f(kx)$ and $\phi_\alpha(x+s)$ is the Airy transform of $f(x+s)$. Therefore we can use the Airy transform of the Gaussian function and the translation rule with $s=\frac{p_2(1-r^4)}{\sqrt{2\hbar r^2(1+r^4)}}$ followed by the scaling rule with $k=\sqrt{\frac{1+r^4}{2\hbar r^2}}$ and to get the required Airy transform. 

The resulting Wigner function is 
\begin{widetext}
    \begin{multline}
        W_\text{CPE}(\mathbf{q},\mathbf{p})=\frac{2^{\frac76}re^{\frac{4r^6}{3(1+r^4)^3\gamma^2\hbar}}}{\pi^{\frac32}\hbar^{\frac{13}{6}}\sqrt{1+r^4}\gamma^\frac13}e^{\frac{2r^2(p_1+p_2+(p_1-p_2)r^4))}{(1+r^4)^2\gamma\hbar}}e^{-\frac{4r^4p_2^2+((q_2-q_1+(q_1+q_2)r^4)^2)}{2r^2(1+r^4)\hbar}}\\\text{Ai}\left(\frac{2^\frac23r^4}{(1+r^4)^2\gamma^\frac43\hbar^\frac23}+\frac{2^\frac23(p_1+p_2+(p_1-p_2)r^4))}{(1+r^4)\gamma^\frac13\hbar^\frac23}+\frac{2^\frac23\gamma^\frac23q_1^2}{\hbar^\frac23}\right)\,.
    \end{multline}
\end{widetext}

\end{document}